\documentstyle[aps,prl,eqsecnum,preprint,tighten]{revtex}
\begin{document}
\title {
Charge and spin density wave ordering transitions in strongly correlated
metals}
\author{Subir Sachdev}
% and T. Senthil}
\address{Department of Physics, P.O. Box 208120, Yale University,
New Haven, CT 06520-8120}
\author{Antoine Georges}
\address{Laboratoire de Physique Th\'{e}orique de l'Ecole Normale
Sup\'{e}rieure,
24 rue Lhomond, 75231 Paris Cedex 05, France}
\date{\today}
\maketitle

\begin{abstract}
We study the quantum transition from a strongly correlated metal, with heavy
fermionic quasiparticles, to a metal with
commensurate
charge or spin density wave order. To this end, we introduce and
numerically analyze  a large dimensionality model of
Ising spins, in a transverse field, coupled to two species of fermions; the
analysis borrows heavily from recent progress in the solution of the
Hubbard model in large
dimensions. At low energies, the Ising order parameter fluctuations
are characterized by the critical exponent
$z \nu = 1$, while above an energy scale, $\Gamma$, there is a crossover to
$z\nu = 1/2$
criticality. We show that $\Gamma$ is of the order of the width of the
heavy quasiparticle
band, and can be made arbitrarily small for a correlated metal close to a
Mott-Hubbard
insulator. Therefore, such a correlated metal has a significant
intermediate energy range
of $z\nu=1/2$ behavior, a single particle spectrum with a
narrow quasiparticle band, and well-formed analogs of the lower and upper
Hubbard bands; we suggest that these features are
intimately related in
general.

\end{abstract}
\widetext

\section{Introduction}
\label{intro}
It has become increasingly clear that studies of
magnetic ordering quantum transitions in metallic, fermionic systems will be of
significant utility in understanding the properties of strongly correlated
systems like
the cuprates or the heavy-fermion
%% FOLLOWING LINE CANNOT BE BROKEN BEFORE 80 CHAR
compounds~\cite{hertz,CHN,jinwu,Ch-Sach,Millis,Sokol,rajiv,sss,pines,monthoux,su,im,scs}.  The
reader is referred to a recent comprehensive study by Barzykin and
Pines~\cite{barrecent}
comparing such ideas with experimental data on the cuprates.

We begin by discussing some background before introducing the question
addressed in this paper.
Consider a metal in the vicinity of a transition to a ground state with
spin or charge
density wave order. We denote by $\Psi_{\mu} (x, \tau)$ the order parameter
field as a function of space ($x$) and Matsubara time ($\tau$)---the index
$\mu=1\ldots3$ for
the case of vector spin density wave order and
$\mu=1$ for the case of scalar charge density wave order.
The long distance, long-time, zero temperature ($T$) effective action for
$\Psi_{\mu}$ obtained after integrating out all the fermionic degrees of
freedom takes the
form~\cite{hertz,Millis,pines,monthoux,su,scs}
\begin{equation}
S = \int d^d q d \omega  |\Psi_{\mu} (q, \omega )|^2 ( r + q^2 + \gamma
|\omega| +
\omega^2 / c_0^2 ) + \ldots .
\label{action}
\end{equation}
We have Fourier transformed to wavevector $q$ and Matsubara frequency $\omega$,
and introduced phenomenological constants $r$, $\gamma$ and $c_0^2$;
the ellipsis represent higher order couplings between the $\Psi_{\mu}$. We
are considering
here only the case originally considered by Hertz~\cite{hertz} in
which~\cite{Millis}
the ordering wavevector, $\vec{Q}$, is not a spanning wavevector of the
Fermi surface,
and it is possible to connect at least two points on the Fermi surface by
$\vec{Q}$---for a
spherical Fermi surface this last condition is $|\vec{Q}| < 2 k_F$, where
$k_F$ is the Fermi
wavevector. In the language of a recent paper by one of us, Chubukov and
Sokol~\cite{scs},
we are considering only transitions of type $B$. Under these conditions,
the density of
particle-hole excitations with wavevector $\vec{Q}$ are linear in energy at
small energies,
and this is responsible for the dissipative $|\omega|$ term in $S$. The
$q^2$ and
$\omega^2$~\cite{pines,monthoux,su,scs}
terms are simply regular terms controlling spatial and dynamic fluctuations.
Alternatively,
the $\omega^2$ term may be viewed as the simplest addition that makes the order
parameter correlator have the correct $1/\omega^2$ decay at large frequencies.

In mean-field theory, $S$ undergoes an ordering transition when the
Landau-like parameter
$r=0$ ---hence $r>0$ measures the distance of the metal from the ordered
state (we caution
that an action like $S$ is {\em not\/} appropriate for $r<0$~\cite{scs}).
For the usual reasons, the mean-field transition has a correlation length
exponent $\nu=1/2$.
At low energies, the $|\omega|$ is the most important term controlling the
energy
scale---comparing this with the $q^2$ term we get a mean-field
value for the dynamic critical exponent of $z=2$~\cite{hertz}. Later we
will consider the
solution of a model for such a transition in the limit of large
dimensionality, where
$z$ and $\nu$ are not separately defined, but the energy scale exponent $z
\nu$ is.
Hence we prefer to quote the value of $z\nu$ which is $z\nu=1$ for the
present mean-field
theory. It has been argued on theoretical and
phenomenological grounds~\cite{Sokol:LANL,Sokol,pines,monthoux,su,scs}
that, under suitable
conditions, the nearly-ordered metallic state should exhibit a crossover at
higher frequency or
temperature scales to region with the effective exponent $z=1$, or in
mean-field theory,
$z\nu=1/2$ (at the $z=1$ fixed point we still have $\nu=1/2$ in large $d$
mean-field theory,
but  $\nu
\approx 0.7$  in $d=2$). It has also been suggested that this $z=1$
behavior is observed in the
cuprates not only at very low doping~\cite{Ch-Sach} but also at moderate
dopings~\cite{Sokol,barrecent}. In a simple mean-field analysis of $S$, the
energy/temperature
scale above which
$z\nu=1/2$ behavior appears is of order
$\gamma c_0^2$, as that is when the $\omega^2$ term starts becoming more
important than the
$|\omega|$ term.

The question we address in this paper is the following: under what
conditions does this
crossover from $z\nu=1$ to $z\nu=1/2$ actually occur ? In other words, is
there a reasonable
scenario in which the crossover scale $\sim \gamma c_0^2$ is significantly
smaller than
all other
higher energy cutoffs of the critical behavior, so that $z\nu=1/2$ behavior
is clearly
observable ? At sufficiently high energies the behavior of the system must
become
dominated by lattice scale cutoffs, and hence non-universal, and it is
therefore important to
have the cutoff $\sim \gamma c_0^2$ be lower than such cutoffs.

We will answer this question here using the solution of a large
dimensionality model of spinless
fermions which has a charge density wave-like order parameter with a $Z_2$
symmetry
(the solution uses recent progress in understanding the Hubbard model in large
dimensions~\cite{metzner,georgkot,jarrell,georges,kotlett}). Our model
exhibits a transition between metallic states with and without a mean value
for the $Z_2$ order
parameter.
We will show that there is indeed a reasonable scenario in which the
vicinity of this transition
exhibits
$z\nu=1/2$ behavior over a  significant intermediate energy scale.
Although we will explicitly display results only for a specific model,
our arguments are quite generic; we expect that similar results apply to
other models of
ordering transitions in fermionic systems in infinite dimensions, including
those with
order parameters with a larger symmetry {\em e.g.\/} the $O(3)$ symmetry in
spin density wave
transitions.

We now turn to a qualitative discussion of the region of parameter space where
such behavior occurs. Rather than introducing the infinite-dimensional
model at this point,
let us discuss a more familiar example in two dimensions; this example has
the virtue of having
phases closely analogous to all of the phases of the infinite-dimensional
model (and perhaps
more). This will make the physical meaning of the phases transparent,
allowing us, then, to
jump directly to the phase diagram of the infinite-dimensional model.

Consider spin-1/2 electrons moving on a
square lattice
with first ($t_1$) and second ($t_2$) neighbor hopping matrix elements and
short
range repulsive interactions of scale $U$ which prefer spin density wave
ordering
at the wavevector $\vec{Q} = (\pi , \pi)$ (See {\em e.g.\/}~\cite{scs}).
We need second neighbor hopping to avoid effects from nested Fermi surfaces.
 On general grounds we anticipate 4 distinct ground states for this model
at half-filling:\\
(A) Metal --- this is a Fermi liquid whose Fermi surface crosses the
boundary of the magnetic
Brillouin zone. Such a Fermi surface is not nested and is also of the type
in which
spin fluctuations are described by an action like $S$.
\\
(B) Metal with spin density wave order --- the magnetic order causes gaps
over portions of
the Fermi surface, but the system remains metallic.\\
Some details of the evolution of the Fermi surface and critical properties
of the
transition between phases
(A) and (B) were discussed in Ref~\cite{scs}. At very large values of $U$ we
can
have two additional insulating phases \\
(C) Insulator with N\'{e}el order --- all charged excitations are now gapped.
The appropriate model for spin excitations is the spin-1/2 Heisenberg spin
model
with first and second neighbor exchange---the so-called $J_1$-$J_2$ model.\\
(D) Insulating quantum paramagnet --- or a Mott-Hubbard insulator. The
$J_1$-$J_2$ model can
have a ground state in which the magnetic order is
short-ranged~\cite{j1j2,peierls}. This
state can also have spin Peierls order~\cite{peierls}, but this is
incidental to our
considerations here.\\
Determining, in this two-dimensional model, the topology of the phase
diagram with phases A, B,
C, D, and possibly others, the nature of the
phase transitions, and of the possible multicritical points, is a
formidable problem
which we shall not address here.

Turning then to the large dimensionality model to be introduced later in
this paper,
we show in Fig~\ref{phasediag} its phase diagram.
The phases are
closely analogous to phases A, B, C, and D above and are labeled as such in
the figure.
The model has a charge density wave-like $Z_2$ order parameter $\left\langle
\tau_z \right\rangle$, but its role is analogous to the N\'{e}el order
parameter
above.

We are now ready to state the main result of this paper. We are interested in
critical properties of the transition from phase A to phase B. Consider the
region of phase A which is close to the Mott-Hubbard insulator D.
We identify this region as a `strongly correlated metal', and has been shaded
as such in the Fig~\ref{phasediag}. In such a
metal~\cite{georges,kotlett,rozen,moeller}
the fermionic quasiparticles are quite heavy and form a narrow band of width
$\Gamma$ around the chemical potential. A large part of the single particle
spectral weight
is in the higher energy lower and upper Hubbard bands.
Now consider the ordering transition from the strongly correlated metal to
the metallic, ordered, phase B; in other words, the transition from
A to B, not too far from the multicritical point M.
We shall show that in the large dimensionality limit, the order parameter
fluctuations are
characterized by
$z\nu =1$ only at energy scales below
$\Gamma$, while $z\nu =1/2$ behavior takes over at the intermediate
energy scales between
$\Gamma$ and
$\sqrt{\Gamma U}$. For all energies below $\sqrt{\Gamma U}$, and in finite,
but large,
dimensions, the action
$S$ in Eqn. (\ref{action}) is an excellent starting point for describing
order parameter fluctuations and the crossover between $z\nu=1$ and
$z\nu=1/2$ criticality,
as examined in Ref~\cite{scs}. As $\Gamma$ can
be made small by moving towards phase D, there is a significant window of
intermediate energies with $z\nu=1/2$
order parameter fluctuations.
(For extremely small values of $\Gamma$, it may also be necessary to
consider the
energy range between $\sqrt{\Gamma U}$ and $U$---this regime also has
$z\nu=1/2$
criticality, but with a significant non-universal renormalization of the
parameter
$c_0$ in $S$ at the scale $\sqrt{\Gamma U}$).
These results suggest
an intimate relationship between a low energy crossover to $z\nu=1/2$
behavior, heavy
quasiparticle bands, and the removal of spectral weight to the lower and upper
Hubbard bands---we conjecture that this relationship
is more generally valid.

Finally, we mention that a recent study~\cite{3he} of liquid ${}^3$He
considers a large
dimensionality model whose phase diagram has many similarities to
Fig~\ref{phasediag}, and
which also appeals to the proximity to a  multicritical point closely
anologous to M.

The infinite dimensional model will be introduced in Section~\ref{model}
and its ground states
discussed in Section~\ref{ground}. The main conclusions will be reiterated in
Section~\ref{conc}.

\section{The Model}
\label{model}
We motivate our model from the $d=2$ spin-1/2 model considered in
Section~\ref{intro}.
In the vicinity of the transition between phases A and B, we need only
focus on the
spin density wave order parameter and portions of the fermi surface which
are close to
pairs of points that can be
connected by the ordering wave-vector $\vec{Q}$. Let $\vec{k}_1$ and
$\vec{k}_2 = \vec{k}_1 + \vec{Q}$ be one such pair of points; in general,
there will
be other pairs of points, usually related to $\vec{k}_1$, $\vec{k}_2$ by
symmetry operations
of the square lattice---these can be treated in a similar manner and are
not considered
explicitly. We introduce two species of fermions $c_{1\alpha}$, $c_{2\alpha}$
($\alpha = \uparrow, \downarrow$ is the spin index) representing Fourier
components of the
electron in the vicinity of $\vec{k}_1$ and $\vec{k}_2$. The low energy
effective action
for the vicinity of the boundary between $A$ and $B$ can be written
as~\cite{scs}
\begin{eqnarray}
\tilde{S} = \int d^d x d\tau &&\left[
(\vec{\nabla} \Psi_{\mu} )^2 + (\partial_{\tau} \Psi_{\mu} )^2 /\tilde{c}_0^2
+ \tilde{r} \Psi_{\mu}^2 + c_{1\alpha}^{\dagger} ( \partial_{\tau}
- \vec{v}_1 \cdot \vec{\nabla} ) c_{1 \alpha}
+ c_{2\alpha}^{\dagger} ( \partial_{\tau}
- \vec{v}_2 \cdot \vec{\nabla} ) c_{2 \alpha} \right.
\nonumber \\
&&~~~~~~~~~~~~~~~~~~~~~~~~~~~~~~~~~~~~~~~~~~~~~~~~~~\left.- \lambda \Psi_{\mu}
(
c_{1\alpha}^{\dagger} \sigma_{\alpha\beta}^{\mu} c_{2 \beta} + \mbox{H.c.})
\right],
\end{eqnarray}
where $\vec{v}_1$, $\vec{v}_2$ are the Fermi velocities at $\vec{k}_1$,
$\vec{k}_2$,
and $\sigma^{\mu}$ are the Pauli matrices. After integrating out the fermions,
the effective action for $\Psi_{\mu}$ should be of the form (\ref{action}).
The dissipative
term is of the form $|\omega|$ as long as $\vec{v}_1$ and $\vec{v}_2$ are
not anti-parallel
{\em i.e.\/} there is no nesting.

We now abstract from $\tilde{S}$ the essential ingredients for a simple
model of the ordering
transition in large dimensions. First, we reduce the symmetry of the order
parameter
from $O(3)$ down to $Z_2$ (the symmetry of a charge density wave order
parameter) by
discarding the spin index
$\alpha$. Then, we replace the scalar field $\Psi$ by an Ising spin
$\tau_z$---its `kinetic'
term then becomes a transverse field, $g$. (For the model with $O(3)$
symmetry, the Ising
spins in a tranverse field would be replaced by $O(3)$ quantum rotors).
Finally, we place the system
on a lattice (not the same lattice for which
$\tilde{S}$ was the continuum limit) and write down the Hamiltonian
\begin{eqnarray}
H = -g \sum_{i} \tau_{xi} && - \frac{1}{Z} \sum_{i>j} J_{ij} \tau_{zi}
\tau_{zj}
-\mu \sum_{i} ( c_{1i}^{\dagger} c_{1i} + c_{2i}^{\dagger} c_{2i} )
- \frac{1}{\sqrt{Z}} \sum_{ij} ( t_{1ij} c_{1i}^{\dagger} c_{1j}
+ t_{2ij} c_{2i}^{\dagger} c_{2j} ) \nonumber \\
&&~~~~~~~~~~~~~~~~~~~~~~~~~~~~~~~~-\lambda \sum_{i} \tau_{zi}
(c_{1i}^{\dagger} c_{2i} + c_{2i}^{\dagger} c_{1i} ).
\label{hamiltonian}
\end{eqnarray}
We have introduced the two Pauli matrices $\tau_z$, $\tau_x$,
an Ising spin exchange $J_{ij}$, a chemical potential $\mu$,
hopping matrix elements
$t_{1ij}$ and $t_{2ij}$ and a fermion-Ising coupling $\lambda$. The
parameter $Z$ has different meanings depending upon the particular large
dimensionality
limit chosen~\cite{metzner,georges,kotlett}---for a regular $d$ dimensional
lattice, $Z=d$; for
the Bethe lattice,
$Z$ is the co-ordination number; and for a fully connected cluster with
random hopping
$t_{1ij}$, $t_{2ij}$ between all pairs of site, $Z$ is the total number of
sites.
The Hamiltonian $H$ contains all low order terms consistent with the $Z_2$
symmetry
\begin{equation}
\tau_z \rightarrow \tau_x \tau_z \tau_x = -\tau_z~~~;~~~\tau_x \rightarrow
\tau_x \tau_x \tau_x = \tau_x~~~;~~~c_1 \rightarrow c_1~~~;~~~c_2
\rightarrow - c_2~~~.
\label{symmetry}
\end{equation}
In the context of the original lattice model of electrons, this $Z_2$
symmetry is simply
the sublattice interchange symmetry which is broken by a charge density
wave ordered state.

We will study in this paper the $Z=\infty$ limit of $H$. Further, we will
focus only
on solutions of the mean field equations in which all sites are equivalent.
This
amounts to neglecting possibilities in which there is a staggered ordering
of the Ising spins
$\tau_z$. Such solutions do occur, and are often lower in energy than the
ones we
consider~\cite{georges,rozen,laloux}; however we are not interested in them
on physical
grounds. One can also appeal to the fully connected random clusters for which
the spatial uniform solutions are expected to be true ground
states~\cite{me,georges,rozen,laloux}.

We will make an additional, final, simplification. We will work with models in
which $t_{1ij} = t_{2ij}$. Although there is no symmetry enforcing this
equality, in
finite-dimensional, regular lattice models such an assumption would
drastically change the physics of
the transition. This is because, as will become clear below, for $t_{1ij} =
t_{2ij}$ there are
additional conserved charges which have a significant effect on the
critical theory~\cite{hertz}.
However at
$Z=\infty$, all of the non-trivial critical behavior is in local
correlators which are
expected to behave in a similar manner for $t_{1ij} = t_{2ij}$ and $t_{1ij}
\neq t_{2ij}$. The numerical analysis required to solve the mean-field
equations
becomes much simpler at $t_{1ij} = t_{2ij}$.

We conclude this section by writing down the mean-field equations.
The equations take their simplest form after performing a rotation of the
fermionic
fields
\begin{equation}
c_a = \frac{1}{\sqrt{2}} ( c_1 + c_2 ) ~~~~~~~~~~~~~~~~~~~
c_b = \frac{1}{\sqrt{2}} ( c_1 - c_2 ).
\end{equation}
Under the $Z_2$ symmetry (\ref{symmetry}) we have $c_a \rightarrow c_b$ and
$c_b \rightarrow
c_a$; this makes it clear that we can simply think of the $a$ and $b$
fermions as
moving predominantly on the two sublattices of the model underlying
$\tilde{S}$.
For $Z=\infty$,
the model is mapped onto a single-site mean-field hamiltonian supplemented
by a self-consistency condition. Following
Refs~\cite{georgkot,georges,kotlett}, for $t_{1ij} =
t_{2ij}=t$ for nearest neighbors on the Bethe lattice, or for the fully
connected cluster with
$t_{1ij} = t_{2ij}$ random but $J_{ij}$ uniform,
the single site mean-field Hamiltonian is
\begin{eqnarray}
H_{MF} = - g \tau_x - \mu && ( c_a^{\dagger} c_a + c_b^{\dagger} c_b ) - J
m \tau_z - \lambda
\tau_z (c_a^{\dagger} c_a - c_{b}^{\dagger} c_b )  \nonumber \\
&& + \sum_{k} \left( \epsilon_{ak} f_{ak}^{\dagger} f_{ak} +
\epsilon_{bk} f_{bk}^{\dagger} f_{bk} - V_{ak} c_a^{\dagger} f_{ak} -
\mbox{H.c.}
- V_{bk} c_b^{\dagger} f_{bk} - \mbox{H.c.} \right) .
\label{hmf}
\end{eqnarray}
We have introduced fermions $f_a$, $f_b$ to model the environment of the site
of
interest.
The couplings of these fermions and the parameter $m$ are determined
by the self-consistency
conditions~\cite{georgkot,kotlett,georges,rozen}
\begin{eqnarray}
m &=& \left\langle \tau_z \right\rangle \nonumber \\
t^2 G_a (\omega ) &=& \sum_{k} \frac{|V_{ak}|^2}{i \omega - \epsilon_{ak}}
\nonumber \\
t^2 G_b (\omega ) &=& \sum_{k} \frac{|V_{bk}|^2}{i \omega - \epsilon_{bk}} ,
\label{selfcon}
\end{eqnarray}
where $G_a$ is the Fourier transform of the $a$ fermion Greens function
$- \left\langle c_a (0) c_a^{\dagger} (\tau ) \right\rangle$ and likewise
for $G_b$.

Note that the total number of $a$ fermions and $b$ fermions is separately
conserved.
This is a consequence of the choice $t_{1ij} = t_{2ij}$; we reiterate that
while such a choice
and the additional conserved quantities would be dangerous in the finite
dimensional theory,
it's effect on local correlators in the $Z=\infty$ limit is expected to be
innocuous.

\section{Ground States of the Model}
\label{ground}

This section will consider solution of the $Z=\infty$ model $H_{MF}$ in Eqn
(\ref{hmf})
along with the self consistency conditions (\ref{selfcon}).
We will consider subsection~\ref{atomic} the exact solution in the atomic
limit $t=0$, $J=0$, followed
by a numerical solution of the $t\neq 0$, $J=0$ case in
subsection~\ref{Jeq0}. The most general
$t\neq=0$, $J> 0$ case will be considered in subsection~\ref{Jgt0}.

\subsection{Atomic Limit, $t=0$, $J=0$}
\label{atomic}
All of the sites are now independent. Each site can have either 0, 1, or 2
fermions
and the eigenenergies of $H$ can be easily determined. They are
\begin{eqnarray}
E_0 &=& \pm g~~~~~,~~~~~\mbox{2 states} \nonumber \\
E_1 &=& -\mu \pm \sqrt{\lambda^2 + g^2}~~~~~,~~~~~\mbox{both doubly
degenerate, 4 states}
\nonumber \\
E_2 &=& -2\mu \pm g~~~~~,~~~~~\mbox{2 states}.
\label{e123}
\end{eqnarray}
A key observation is that if we take the lower eigenvalue for each particle
number,
the states map exactly onto those of the Hubbard model with $U/2 =
\sqrt{\lambda^2 + g^2}
-g$. Moving away from the atomic limit, the mapping of the low energy states
to the Hubbard model
will continue to hold as long as $t,J < g$. This mapping will be very
useful to us
in the subsequent discussion.

\subsection{$J=0$}
\label{Jeq0}

We will study the ground states at fixed values of $\lambda$ and $t$ as a
function
of $g$. This corresponds to the $y$-axis in Fig~\ref{phasediag} which
presents results at
$t=1$ and $\lambda=2.5$.

First consider the limits of large and small $g$.

For large $g$, the Ising spin flips
 rapidly between its up and down states as a consequence of the $g \tau_x$
term.
The value of $\tau_z$ averages out to zero, and the fermions effectively do
not see the Ising
spin. The fermion spectral function $\rho_{a,b} ( \Omega ) = \mbox{Im}
G_{a,b} ( \Omega)$ is then simply the semi-circular density of states of
free fermions:
\begin{equation}
\rho_a ( \Omega ) = \rho_b (\Omega) = \frac{1}{2 t^2} \sqrt{4t^2 -
(\Omega-\mu)^2 }~~~;~~~g
\rightarrow ,
\infty
\end{equation}
for $|\Omega-\mu| < 2t$ and zero otherwise.

The behavior at small $g$ is a little
more subtle. The Ising spin now fluctuates slowly between its up and down
states and
the fermions have plenty of time to respond to its instantaneous
orientation: this
yields fermion bands centered around $\pm \lambda$.
As the Ising spin is equally likely to be up or down, the fermion spectral
function
is an equal superposition of the two possibilities
\begin{equation}
\rho_a ( \Omega ) = \rho_b (\Omega) \approx \frac{1}{4t^2} \left[
\sqrt{4t^2 - (\Omega-\lambda-\mu)^2} + \sqrt{4t^2 - (\Omega+\lambda-\mu)^2}
\right],
{}~~~g~\mbox{small}
\label{gsmall}
\end{equation}
where it is assumed that the square roots vanish when their arguments are
negative. For $\lambda > 2 t $ there is a window of energies where
(\ref{gsmall})
predicts a gap; in reality, we will only have a pseudogap whose origin is
related to the phenomenology discussed by
Kampf and Schrieffer~\cite{shadow}. For $g > t$ (see Section~\ref{atomic})
it is more appropriate to think of the
pseudogap as arising from the formation of the
 `upper and lower Hubbard bands'; there is, however,
no fundamental distinction between the two mechanisms and they continuously
evolve into each
other.

One might also wonder why, for small $g$, we do not expect the $Z_2$
symmetry to be broken,
thus causing the Ising spin to pick a definite orientation and $\rho_a \neq
\rho_b$.
This, however, does not happen for the same reason that the insulating
solution of the Hubbard
model in Refs~\cite{georges,kotlett,rozen} is not ferromagnetic. One can
test instability
towards `ferromagnetism' by measuring the response to an external
field---even though
the local susceptibility is infinite in the insulating solution, there is
an internal
field from the neighboring fermions which causes the net local effective
field to
vanish~\cite{rozen}, and the uniform susceptibility remains finite; the reader
is referred to the
discussion in Refs~\cite{rozen,revmod} for more discussion on this
important point in the context of
the Hubbard model.

We have numerically studied the crossover from small to large $g$
by solving the Eqns (\ref{selfcon}) using the exact diagonalization
method~\cite{caffarel,si}. We used as many as 8 sites each for the $a$ and $b$
fermions and worked at half-filling. The results were very similar to those
obtained in the Hubbard
model~\cite{georges,kotlett,rozen}. We obtained two classes of solutions, a
metallic solution
present for $g > g_{c2}$ and an insulating solution present for $g< g_{c1}$
with
$g_{c1} > g_{c2}$. For $\lambda=2.5$ and $t=1$ (the parameters used in
Fig~\ref{phasediag}),
the energies of the two solutions crossed each other at a value of $g
\approx 0.7$ which was
quite close to
$g=g_{c2}$. We are therefore uncertain as to whether the metal-insulator
transition was first
or second order. However this issue is really peripheral to what we are
interested in here, and
so we did not do the necessary investigation of numerical fine structure to
resolve it.

What we do care about is the behavior of the metallic solution for $g$
close to but greater
than $g_{c2}$---{\em i.e\/} the shaded region of the strongly correlated
metal in
Fig.~\ref{phasediag}. In this region, it is appropriate to use
the continuous transition at $g=g_{c2}$ to obtain
a scaling description of the
response functions. There is a low energy scale, which we denote by
$\Gamma$, which
vanishes as $g$ approaches $g_{c2}$ as~\cite{rozen}
\begin{equation}
\Gamma \sim g - g_{c2}.
\end{equation}
The scale $\Gamma$ controls the width of the quasiparticle band ($\sim
\Gamma$) or the
effective mass ($\sim 1/\Gamma$). Our main interest here is in the behavior
of dynamic
correlations of the order parameter $\tau_z$. Its scaling properties can be
deduced
from the elegant critical theory of the Mott transition provided recently by
Moeller {\em et. al.\/}~\cite{moeller}.
(Our model $H$ has only a $Z_2$ symmetry and so the critical theory will be
a self-consistent
Kondo-like model~\cite{moeller} but with a planar anisotropy in the
exchange constants).
For the response, $\chi_{loc}$, to a local field
coupling to
$\tau_z$
\begin{equation}
\chi_{loc} ( \omega ) = \int d\tau e^{-i\omega \tau} \left \langle
\tau_z (0) \tau_z (\tau ) \right \rangle,
\end{equation}
we have~\cite{moeller}:
\begin{equation}
\chi_{loc} ( \omega ) = \frac{1}{\Gamma} \phi \left( \frac{\omega}{\Gamma}
\right)
+X_1 - X_2 \omega^2 + \ldots~~~~~~~~|\omega| < \Lambda ,
\label{chiloc}
\end{equation}
with $\phi$ a universal scaling function, $\Lambda$ is an upper cutoff of order
$\lambda$ or $g$, and constants $X_1$, $X_2$ contain the corrections to
scaling contributions of the
higher energy excitations associated with lower and upper Hubbard bands.
The latter excitations
have an energy $\sim \Lambda$ and hence their contribution can be expanded
in a smooth power
series in even powers of $\omega$. On dimensional grounds we expect $X_1
\sim 1/\Lambda$
and $X_2 \sim 1/\Lambda^3$.

We show in Fig~\ref{scaling} a test of the scaling form (\ref{chiloc})
using the numerically computed
value of $\chi_{loc}$ at four values of $g > g_{c2}$---the collapse of the
data onto a single
scaling curve is quite reasonable. The scaling function ${\phi}$ can be chosen
such that $\phi (0) = 1$ and must
satisfy the asymptotic limits
\begin{equation}
\phi ( x ) = \left\{
\begin{array}{ll} 1 - \tilde{c}_1 |x| & \mbox{for $|x| \rightarrow 0$} \\
\tilde{c}_2 / x^2 & \mbox{for $|x| \rightarrow \infty$}
\end{array}
\right. ,
\label{asymptote}
\end{equation}
for some positive constants $\tilde{c}_1, \tilde{c}_2$. The non-analytic
$|x|$ behavior at small
frequencies is a consequence of the damping of spin excitations from the
finite density of states at
the Fermi level in the metal, while the $x^2$ behavior at large frequencies
follows simply from the
spectral representation of the response functions in the critical theory.
These asymptotic
limits suggest the simple interpolation form
\begin{equation}
\phi (x) \approx \frac{1}{1 + c_1 |x| + c_2 x^2},
\label{interpolate}
\end{equation}
which satisfies (\ref{asymptote}). We also show in Fig.~\ref{scaling} a fit
of the measured
scaling functions to (\ref{interpolate}). It is apparent that the
interpolation form works
rather well, and that the accuracy of the numerical results is not
sufficient to distinguish
between the true ${\phi}$ and the approximate form (\ref{interpolate}). For
the parameters
chosen, the best fit values were $c_1 \approx 0.07$ and $c_2 = 0.32$; the
value of
$c_2$ is quite reliable, but the same cannot be said of $c_1$---
slightly different choices in the fitting process
gave values of $c_1$ differing by a factor of 2, but
only a few percent changes in $c_2$. However, $c_1$ always remained
significantly smaller than
$c_2$---indeed,
the linear $|\omega|$ dependence of $\phi$ is barely visible in
Fig~\ref{scaling}. We had
to compute $\chi_{loc}$ at fairly large values of $g$ ($g > 2.0$), lying
well within phase A, before
the metallic
$|\omega|$ behavior was clearly evident.

The scaling behavior of the $q=0$ susceptibility, $\chi_{q=0} ( \omega )$,
which
determines the dynamic response of the $\tau_z$ to a spatially-uniform but
time-dependent
field acting on the $\tau_z$, is somewhat more complicated (Note that since
$\tau_z$ mimics
the density wave order parameter at a wavevector $\vec{Q}$ in the physical
system,
the $q=0$ susceptibility of $H$ is really the physical staggered
susceptibility).
Unlike, $\chi_{loc}$ this susceptibility is finite at the transition at
$g=g_{c2}$~\cite{georges,kotlett,rozen};
as a result the interesting frequency dependence of $\chi_{q=0}$ is really in a
correction to scaling contribution. With the limited accuracy of the
numerical method
we are using, it would be quite difficult to obtain the scaling results
from the data.
We therefore restrict ourselves here to a qualitative discussion, which
will be sufficient to
extract the physics we are interested in. Because of the similarity in the
low energy
structure of $H$ to the Hubbard model (in particular, the two-fold
degeneracy of the lowest
energy state $E_1$ in (\ref{e123}) in atomic limit), we expect that
$\chi_{q=0}$
will behave in a manner similar to the uniform spin susceptibility near the
Mott transition. From arguments similar to those in Ref~\cite{rozen} we may
then
deduce that
$\chi_{q=0}^{-1}$ behaves roughly like
\begin{equation}
\chi_{q=0}^{-1} \sim X + c_3 \chi_{loc}^{-1},
\label{chiqz}
\end{equation}
where the constant $X \sim \Lambda$ and $c_3$ is a dimensionless constant
of order unity (on
the hypercubic lattice, the above equation actually requires $t_{1ij} \neq
t_{2ij}$).  Combining
(\ref{chiloc}), (\ref{asymptote}) and (\ref{chiqz}), we can now deduce the
frequency dependency of
the uniform dynamic susceptibility:
\begin{equation}
\chi_{q=0}^{-1} ( \omega ) \sim X +
\left\{ \begin{array}{ll}
c_3 (\Gamma + c_1 |\omega| + c_2 \omega^2 / \Gamma) &~~~~\mbox{for $ |\omega| <
\sqrt{\Gamma\Lambda}$} \\
c_3 (1/X_1 + X_2 \omega^2 / X_1^2 + \ldots) &~~~~\mbox{for
$\sqrt{\Gamma\Lambda} <
|\omega| < \Lambda$}
\end{array}\right.  .
\label{zeroJ}
\end{equation}
The first of these results in the energy range $|\omega| < \sqrt{\Gamma
\Lambda}$
will be used below to reach the main conclusions of this paper.

\subsection{$J > 0$}
\label{Jgt0}
The exchange coupling $J$ plays a very simple role in the $Z=\infty$
mean-field theory.
It is apparent from Eqns (\ref{selfcon}) that in phases with $m =
\left\langle \tau_z
\right\rangle = 0$ the single particle Green's functions are in fact
independent of $J$.
The results at $J=0$ thus continue to apply for a finite range of values of
$J$.

There is however a change in the $\tau_z$ correlation functions. It is not
difficult to
show, using an argument quite similar to that in the original Curie-Weiss
mean field theory,
that
\begin{equation}
\left. \chi_{q=0}^{-1} ( \omega ) \right|_{J\neq 0} =
\left. \chi_{q=0}^{-1} ( \omega ) \right|_{J = 0} - J .
\label{nonzeroJ}
\end{equation}
The finite $J$ uniform susceptibility therefore diverges at $J= \left.
\chi_{q=0}^{-1} ( \omega
) \right|_{J = 0}$, which is the point we have onset of a nonzero
$\left\langle \tau_z
\right\rangle$.
The phase boundary between phases A and B and between D and C in
Fig~\ref{phasediag} was
determined in this manner. The phase boundary between phases C and D
requires computations
in the phase with $\left\langle \tau_z
\right\rangle\neq 0$; as this phase boundary is of no interest to us here,
we did not carry out the rather involved computations required---the
boundary between C
and D, shown in  Fig~\ref{phasediag} is just an educated guess.

Fig~\ref{phasediag} also shows a multicritical point M, which is the point
where all four
phases would meet if the metal-insulator transition was second-order. The
correlation
functions on the $\left\langle \tau_z \right\rangle =0$ side of this
critical point are however
simply related to those at the $g=g_{c2}$, $J=0$ critical point discussed
in Section~\ref{Jeq0}: the
single particle correlators are unchanged, while the relationship in the
spin susceptibility follows
from (\ref{nonzeroJ}).

The central interest of this paper is in the nature of the order parameter
fluctuations
near the phase boundary between A and B, at a point not too far from M
(Fig~\ref{phasediag}). From (\ref{nonzeroJ}) and (\ref{zeroJ}) we see that the
static
uniform susceptibility
diverges at $J=X$, and the Landau parameter $r = X + \Gamma - J$. The
nature of the dynamic
susceptibility near the phase boundary between A and B also follows from
(\ref{nonzeroJ}) and
(\ref{zeroJ}). From these equations we see that the infinite dimensional
transition
has $z\nu=1$ at the lowest energies; however this behavior is only present for
energies smaller than $\Gamma$, and $z\nu=1/2$ criticality takes over for
larger energy
scales. Observe, however, by comparing the two equations in (\ref{zeroJ}),
that there
is a significant change in the coefficient of the $\omega^2$ term at a scale
$\sqrt{\Gamma\Lambda}$.

We can in fact also make a crude connection between the dynamic
susceptibility computed here
and the finite-dimensional action $S$ for discussed in Section~\ref{intro}.
One perspective on the meaning of the results for dynamic correlators in
the infinite dimensional
model is that these specify the input form of the effective action that
must then be used to
understand fluctuations in finite dimensions. With this point of view, our
results allow us to deduce the following effective action for the order
parameter $\Psi_{\mu}$ in
finite dimensions:
\begin{equation}
S = \int d^d q d \omega  |\Psi_{\mu} (q, \omega )|^2 ( q^2 +
\chi^{-1}_{q=0} ( \omega ) ) + \ldots,
\label{newaction}
\end{equation}
where the $q^2$ has been added on phenomenological grounds and, as in
(\ref{action}), the ellipsis
represent non-linearities not explicitly displayed. From (\ref{zeroJ}) and
(\ref{nonzeroJ}) we see
again that
$r= X + \Gamma - J$, and that the two actions (\ref{action}) and
(\ref{newaction}) have identical
frequency dependencies in the frequency range $|\omega | < \sqrt{\Gamma
\Lambda}$ where
\begin{equation}
S = \int d^d q d \omega  |\Psi_{\mu} (q, \omega )|^2 ( r+ q^2 + c_3 c_1
|\omega| +  c_3 c_2
\omega^2 /
\Gamma) +
\ldots .
\label{newnewaction}
\end{equation}
The insight gained
from the present approach is that we now have an estimate of the scale at which
the crossover from $z\nu=1$ to $z\nu=1/2$ behavior occurs---it is the small
energy $\Gamma$,
as had been claimed earlier in the paper.
The crossover itself is described by the action $S$ in
(\ref{newnewaction}), which is
an excellent approximation for energies below $\sqrt{\Gamma \Lambda}$.
In the higher energy range $\sqrt{\Gamma \Lambda}
< |\omega | < \Lambda$ we should use instead the second result for
$\chi_{q=0}^{-1}$
in (\ref{zeroJ})---this range still has $z\nu=1/2$ behavior, but there is a
non-universal
crossover in the coefficient of the $\omega^2$ term at the boundary
$|\omega | \sim \sqrt{\Gamma
\Lambda}$. So if $\Gamma$ is extremely small, this second regime of $z\nu=1/2$
behavior will be the
most important, and the very low energy regime of $z\nu=1$ behavior will be
unobservable.

Before concluding, we also note that a similar approach can be used to
describe the
ordering transition between the insulating phase D and C. In this case
$\Gamma = 0$
and we are always in the second of the regimes in (\ref{zeroJ})---the
transition therefore
has $z\nu=1/2$ down to the lowest energy scales.

\section{Conclusions}
\label{conc}
This paper has examined a simple, infinite-dimensional model for spin or
charge density
wave ordering transitions in strongly correlated metals. We were interested
in the case, originally
considered by Hertz~\cite{hertz}, in which Fermi surface geometry was such
as to induce a $|\omega|$
damping in effective action for the critical modes of the order parameter.
As a result, the order parameter fluctuations have $z\nu=1$ at the lowest
energy scales.
We examined the conditions under which there is a well-defined crossover at
intermediate energy scales to $z\nu=1/2$ criticality.
We showed that, under suitable conditions, it was indeed possible to have a
low energy scale $\Gamma$ such that the $z\nu =1$ behavior was restricted
to $|\omega| < \Gamma$, and universal $z\nu = 1/2$
behavior appeared in the energy range $\Gamma < |\omega| < \sqrt{\Gamma
\Lambda}$; here $\Lambda$
is an upper cutoff of order the repulsion energy between the
electrons---$\Lambda \sim U$.
For all energies smaller than $\sqrt{\Gamma \Lambda}$ the order parameter
fluctuations in large, but finite, dimensions are expected to be well
described by the action
$S$ in Eqn (\ref{action}) or (\ref{newnewaction}):
a detailed analysis of the finite temperature crossovers and fluctuations
in a $d=2$ system
controlled by $S$ has already been presented in Ref~\cite{scs}.
For frequencies larger than $\sqrt{\Gamma \Lambda}$, the order parameter
fluctuations
still have $z\nu =1/2$---the main caveat to keep in mind is that there is
non-universal change in
the value of
$c_0$  (see (\ref{action}) around $\omega \sim \sqrt{\Gamma \Lambda}$ which
is not describable
by the  simple frequency dependence in $S$.

The key feature of the above scenario is of course the presence of the low
energy scale
$\Gamma$ in a strongly correlated metal. In infinite dimensions, $\Gamma$
vanishes at the
transition to a Mott Hubbard insulator, which we assume (see
Fig~\ref{phasediag}) is in the
vicinity of the strongly correlated metal. In this same
region~\cite{georges,kotlett,rozen},
the single-particle spectral function has a narrow quasiparticle band of
width $\Gamma$,
and additional spectral weight at the energies $\sim \Lambda$ which form the
analogs of the upper and lower Hubbard bands. This window of energy between
the quasiparticle
excitations and the localized Hubbard excitations is directly responsible
for the $z\nu=1/2$
criticality in the order parameter fluctuations.

\acknowledgements
S.S. would like to thank T. Senthil for his interest, probing questions,
and many helpful
remarks on the subject matter of this paper. We also had
beneficial discussions with M.~Caffarel, A.~Chubukov, W.~Krauth,  M.~Rozenberg,
R.~Shankar, Q.~Si, and A.~Sokol.

The research was supported by NSF Grant No. DMR-92-24290.

\begin{figure}
\caption{
Phase diagram of the Hamiltonian $H$~(\protect\ref{hamiltonian}) at
half-fliing in the $Z=\infty$
limit. We have chosen $t=1$ and $\lambda = 2.5$. The metal-insulator
transition between
phase A and D is assumed to be a second-order transition at $g=g_{c2}$.
However,
the phase D continues to be metastable for $g_{c1} < g < g_{c2}$ and the
energy of this state
crosses that of A at a point very close to $g=g_{c2}$. Thus this transition
could be weakly
first-order, in which case $g_{c2}$ and the multicritical point M will be
in a metastable region.
However, the critical theories at $g=g_{c2}$ and $M$ could still be used to
describe the
shaded strongly correlated metal. The Ising order parameter is
$\left\langle \tau_z \right
\rangle$. We are interested mainly in the Ising transition between A and B from
the shaded region of A: in this case the Ising fluctuations have $z \nu =
1$ at low
energies and $z\nu = 1/2$ at intermediate energies. The Ising transition
between D and C
has $z\nu=1/2$ at low and intermediate energies. The phase boundary between
the B and C
is of no real interest to the discussion in this paper, and its position is
based on an
educated guess; positions of other phase boundaries were numerically
determined.
}
\label{phasediag}
\end{figure}
\begin{figure}
\caption{
Scaling plot of the local Ising spin correlation function $\chi_{loc}$ for
$t=1$, $\lambda=2.5$ in phase A of Fig~\protect\ref{phasediag}. The values of
$\Gamma$ with
their respective values of $g$ are: $g=0.75$, $\Gamma=0.0056$;
$g=0.8$, $\Gamma=0.0090$;
$g=0.85$, $\Gamma=0.0128$;
$g=0.0$, $\Gamma=0.0179$. The fit function $\phi$ is defined in Eqn
(\protect\ref{interpolate}),
and the values of the fitting parameters are quoted below it.
}
\label{scaling}
\end{figure}
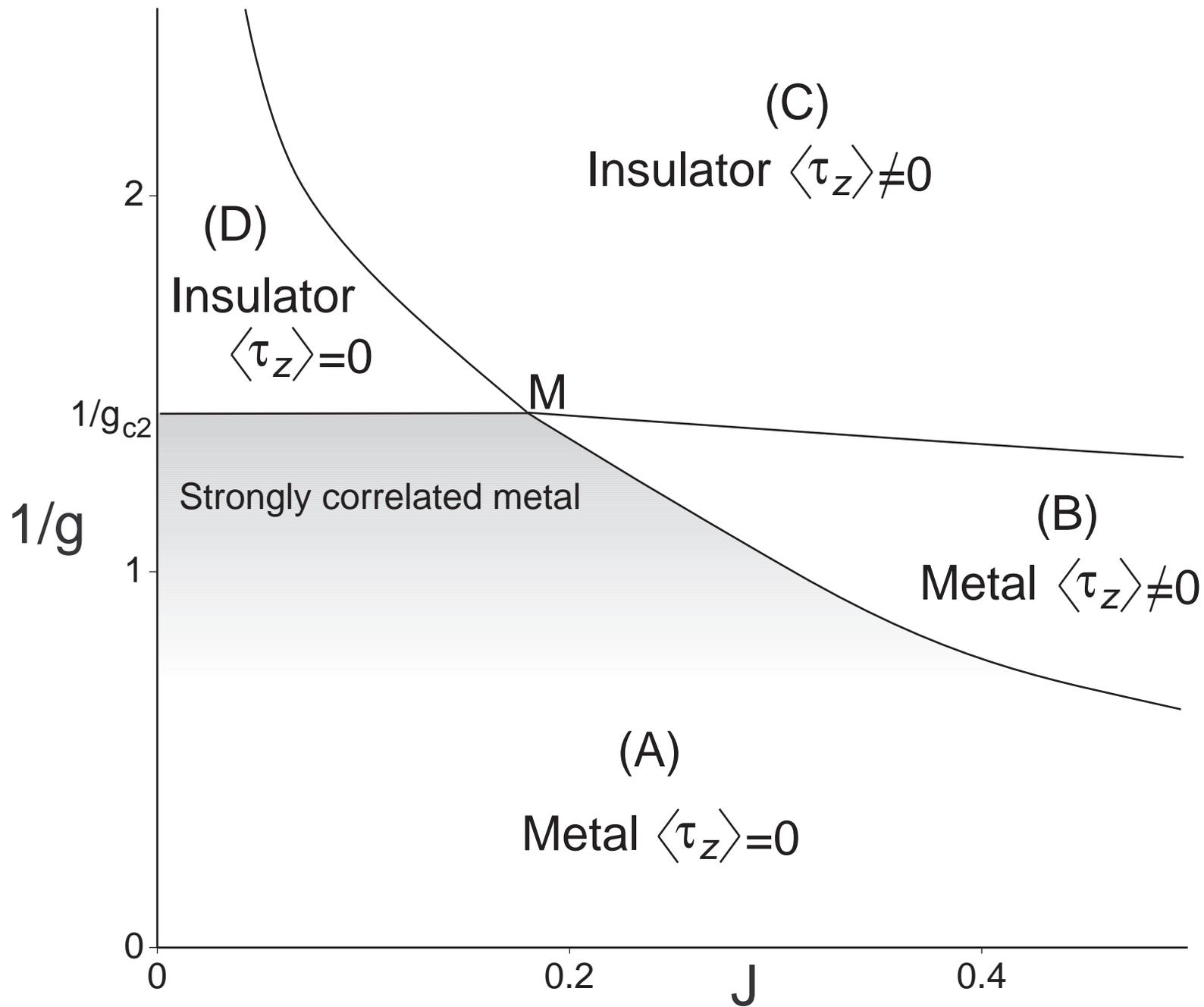
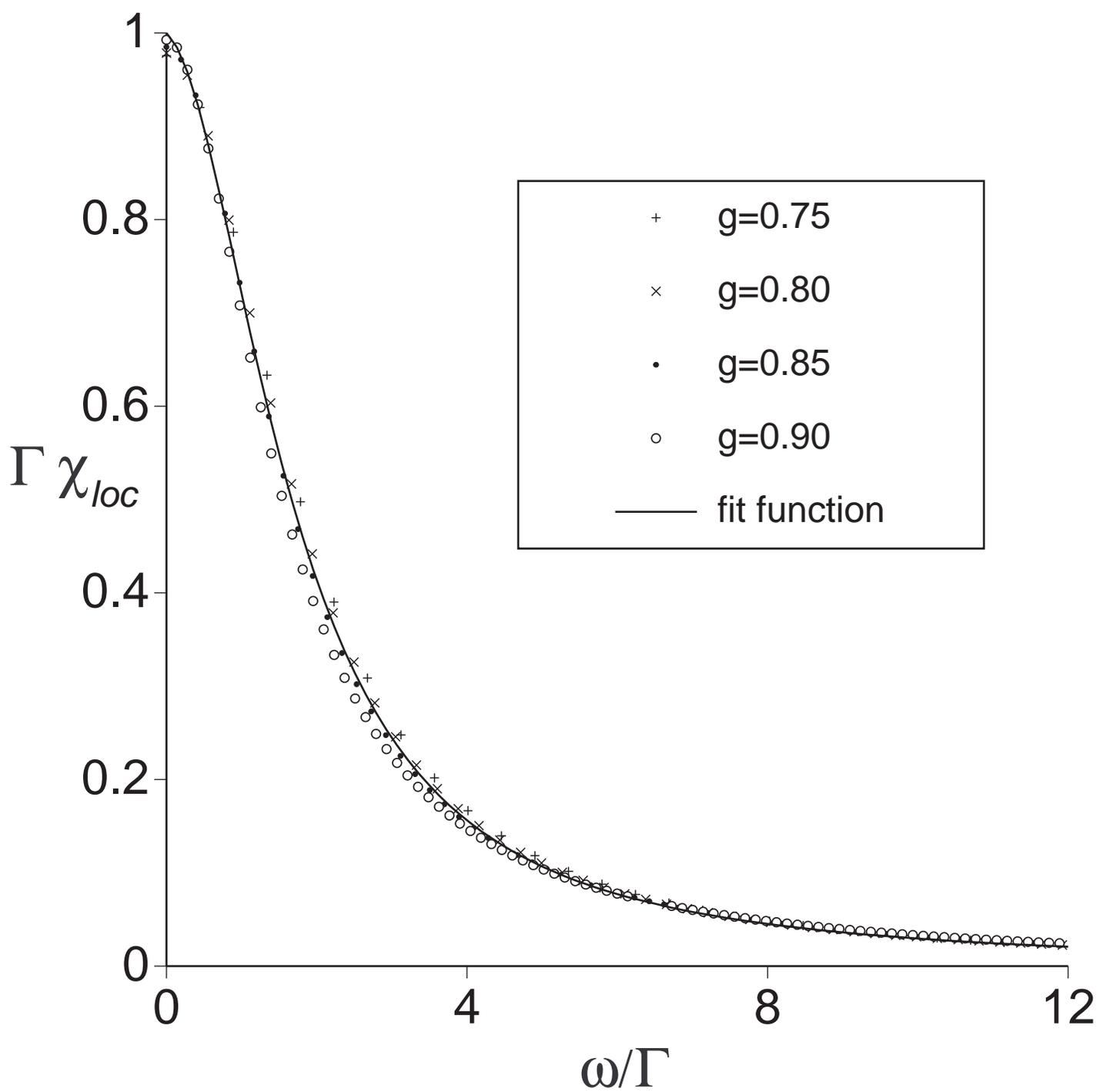
\end{document}